\documentclass[
    prd, twocolumn,
    superscriptaddress, nofootinbib, amsmath, amssymb,
    aps, floatfix
]{revtex4-1}

\usepackage{graphicx,xcolor,setspace}
\usepackage[
    colorlinks=true,
    linkcolor=blue,
    urlcolor=blue,
    citecolor=blue
]{hyperref}

\usepackage[capitalise]{cleveref}
\usepackage{siunitx}
\usepackage{color,epsfig}
\usepackage{ifpdf}
\usepackage{amsmath}
\usepackage{bm}
\usepackage[english]{babel}
\usepackage{amsfonts}%
\usepackage{amssymb}
\usepackage{braket}
\usepackage{hyperref}
\usepackage{enumerate}
\usepackage{cancel}
\usepackage{multirow}
\usepackage{tikz}
\usepackage[compat=1.0.0]{tikz-feynman}

\usepackage{ulem}
\usepackage{array}
\usepackage{ulem,fancyvrb}
\usepackage{xcolor}

\definecolor{nicered}{rgb}{0.7,0.1,0.1}
\definecolor{nicegreen}{rgb}{0.1,0.5,0.1}
\hypersetup{colorlinks,citecolor= nicegreen,linkcolor= nicered}

\usepackage{dcolumn}
\usepackage{bm}
\usepackage{slashed}

\def\comments{true}

\ifx\comments\undefined
	\newcommand{\comment}[1]{}
\else
	\newcommand{\comment}[1]{#1}
\fi

\definecolor{maroon}{cmyk}{0,0.87,0.68,0.32}

\begin{document}

\title{Can Sterile Neutrino Explain Very High Energy Photons from GRB221009A?}

\author{Shu-Yuan Guo}
\email{shyuanguo@ytu.edu.cn}
\affiliation{Department of Physics, Yantai University, Yantai 264005, China}

\author{Maxim Khlopov}
\email{khlopov@apc.in2p3.fr}
\affiliation{Virtual Institute of Astroparticle physics, 75018 Paris, France\\
Institute of Physics, Southern Federal University, Stachki 194, Rostov on Don 344090, Russia\\Center for Cosmoparticle Physics Cosmion, National Research Nuclear University “MEPHI”, 115409 Moscow, Russia}

\author{Lei Wu}
\email{leiwu@njnu.edu.cn}
\affiliation{Department of Physics and Institute of Theoretical Physics, Nanjing Normal University, Nanjing, 210023, China}

\author{Bin Zhu}
\email{zhubin@mail.nankai.edu.cn}
\affiliation{Department of Physics, Yantai University, Yantai 264005, China}

\begin{abstract}
The LHAASO collaboration has reported their observation of very high energy photons ($E^{max}_\gamma \simeq 18$ TeV) from the gamma-ray burst GRB221009A. The sterile neutrino that involves both mixing and transition magnetic moment may be a viable explanation for these high energy photon events. However, we demonstrate that such a solution is strongly disfavored by the cosmic microwave background (CMB) and Big Bang nucleosynthesis (BBN) in the standard cosmology. 
\end{abstract}

\maketitle

\section{Introduction}

The most energetic cosmic explosions generate gamma-ray bursts (GRB), which provide a crucial window into the extreme Universe. An unprecedented bright gamma-ray burst, at $z\simeq 0.15$~\cite{redshift,redshift-2}, was first recorded by the Burst Alert Telescope(BAT)~\cite{BAT} on the Swift satellite recently. It was later confirmed by Fermi Gamma-ray burst Monitor(GBM)~\cite{FERMI} and Fermi-LAT~\cite{LAT-1, LAT-2}, and was documented as GRB221009A. The extremely energetic photon emitted from the burst has been captured by LHAASO~\cite{LHAASO} and Carpet-2~\cite{Carpet-2}. In particular, the KM2A detector on LHAASO has reported the observation of $\sim 5000$ very-high-energy(VHE) photons with the energy up to $18~\rm{TeV}$ in a $\sim 2000$s time window. Such energetic photon is expected to be virtually impossible to travel across a distant way. High energy photons will inevitably be attenuated by the extragalactic background light(EBL)~\cite{Gould:1966pza}, i.e., $\gamma+\gamma_{\rm{~EBL}} \to e^+ +e^-$. To be precise, the survival probability of $18~\rm{TeV}$ photon traveling a distance of $z\simeq 0.15$ to arrive at the Earth is around $10^{-7}$~\cite{Dominguez:2011,Franceschini:2017iwq}. Thus it is astonishing to observe plenty of high-energy photons on Earth, which may indicate new physics beyond the Standard Model (SM). Quite a few ideas have been proposed, such as Lorentz invariance violation~\cite{Baktash:2022gnf, Li:2022vgq, Vardanyan:2022ujc}, dark photon~\cite{Gonzalez:2022opy}, axion-like particles~\cite{Baktash:2022gnf, Carenza:2022kjt, Gonzalez:2022opy, Galanti:2022pbg, Lin:2022ocj, Troitsky:2022xso, Nakagawa:2022wwm, Zhang:2022zbm, Galanti:2022xok,Wang:2023okw}, invisible neutrino decay~\cite{Huang:2022udc}, light scalar decay~\cite{Balaji:2023nbn}, sterile neutrinos~\cite{Cheung:2022luv, Smirnov:2022suv, Brdar:2022rhc} and ultrahigh-energy cosmic-ray acceleration in GRB~\cite{Das:2022gon}.

Among them, the sterile neutrino is motivated by the fact that if there exists mixing or interaction between the active and sterile neutrinos, long traveling without decay could be realized. The active neutrinos produced in the gamma-ray burst will convert into sterile neutrinos, then travel a long distance without scattering with the EBL, after that they decay into active neutrinos and photons, which reach the detectors on Earth. The IceCube has reported various results in searching for high-energy neutrinos, e.g., TeV to PeV. For GRB221009A, IcuCube has performed a track-like muon neutrino events search, the non-observation of such events has set an upper limit on muon neutrino flux, i.e. $E^2 dN_{\nu_\mu}/dE < 3.9 \times 10^{-2}~\rm{GeV}\cdot \rm{cm}^{-2}$ at $90\%$ CL~\cite{ICB}.

In this work, however, we point out that the sterile neutrino proposal may not be proper since it faces strong constraints from astrophysical and cosmological observations. Our result, confirming the notice of~\cite{Smirnov:2022suv, Brdar:2022rhc} that existence of such a sterile neutrino would lead to a non-standard cosmology, excludes the corresponding cosmological scenario for sterile neutrino parameters that are needed to reproduce the LHAASO observations. In particular, the $\rm{keV}$ to $\rm{MeV}$ scale sterile neutrino needed in the previous works would contribute a sizable $\Delta N_{\rm{eff}}$, which is in strong contradiction with constraints from BBN and CMB.

\section{Constraints on sterile neutrino explanation \label{sterile}}

Active neutrinos are produced associated with photons in the gamma-ray burst. They could convert into sterile neutrinos through mixing or dipole interaction. Sterile neutrinos can travel a long distance and convert back into active neutrinos and photons where the EBL attenuation is not prominent. Based on the different ways, one could conclude that the production and decay could be categorized into 1) both through mixing with active neutrino; 2) both through dipole interactions $d_\alpha \overline{\nu_{\alpha L}} \sigma_{\mu \nu} F^{\mu \nu} N$($\alpha=e, \mu, \tau$), and 3) produced through mixing(dipole interaction) while decaying through dipole interaction(mixing). Since there is only the IceCube track-like events limit on hand, we make the following discussions under the assumptions that the mixing is only between muon neutrino and sterile neutrino, and $d_\alpha$ is nonzero just for $d_\mu$. For the first case, mixing and mass of sterile neutrino would totally determine the produced photon flux. According to Refs.~\cite{Brdar:2022rhc, Smirnov:2022suv}, the mixing needed to match the observed number of energetic photons will exceed constraints from oscillation experiments searching for sterile neutrinos~\cite{MINOS:2017cae}. For the second scenario, the production of sterile neutrinos is severely constrained by the dipole strength and suppressed by phase space of three body decay~\cite{Brdar:2022rhc, Cheung:2022luv}. The production through mixing will always be dominant once the mixing exists. 
Thus, an alternative scenario that sterile neutrinos are produced through mixing while decaying through dipole interaction has been studied in~\cite{Brdar:2022rhc}. The expected photon flux is given as
\begin{equation}
     \frac{|U_{\mu 4}|^2 d/\tau}{\gamma/\Gamma-d/\tau} \left(\exp\left[-\frac{d\Gamma}{\gamma}\right] -\exp(-\tau) \right) \frac{dN_{\nu_\mu}}{dE} \rm{Br}_\gamma.
\end{equation}
Here $U_{\mu 4}$ is the particular element of the mixing matrix being referred to. The burst is separated from Earth by a distance of $d$, and $\tau$ denotes the optical depth at a redshift of $z=0.15$. $\Gamma$ stands for the total decay width of sterile neutrinos. $\rm{Br}_\gamma$ is the branching ratio of sterile neutrinos decaying into photons, where $\gamma$ labels the relativity boost factor $\gamma=E_N/m_N$. $dN_{\nu_\mu}/dE$ represents the neutrino flux which takes the aforementioned upper limit from IceCube~\cite{ICB}. One could then get the expected event number of photons to be detected. However, we find that the feasible parameter space is tightly constrained by  astrophysical and cosmological observations.

\noindent{\it Stellar Energy Loss Constraints.}  The sterile neutrino $N$ that we are studying can carry energy when it is produced under thermal conditions in a stellar system. It affects the energy loss, thermal conductivity, and the subsequent time evolution of the stellar population, as has been documented in various studies {(see, e.g., \cite{Raffelt:1994ry, Diaz:2019kim})}. In this way, the neutrino's behavior can significantly impact the development of stellar systems. Plasmon decays are kinematically allowed in the dense charged medium where $m_{\gamma^*}>T_{\mathrm{star}}$
\begin{equation}
\gamma^* \longrightarrow \nu_L+N
\end{equation}
This additional energy loss mechanism affects stellar evolution in terms of the increased fuel burning rate. The energy loss per unit volume is given in~\cite{Vogel:2013raa}
\begin{equation}
Q=\int_0^{\infty} \frac{k^2 d k}{\pi^2} \int_{m_N^2}^{\infty} \frac{d \omega^2}{\pi} \frac{\omega \Gamma_T}{\left(K^2-\omega_p^2\right)^2+\left(\omega \Gamma_T\right)^2} \frac{\omega \Gamma_{\gamma^*}}{e^{\omega / T_\gamma}-1}
\end{equation}
where $K$ is the effective plasmon mass and determined by the plasmon energy $\omega$ and momentum $k$ via $K=\sqrt{\omega^2-k^2}$. $\Gamma_T$ is the Thomson scattering rate, $\Gamma_T=\frac{8 \pi \alpha^2 n_e}{3 m_e^2}$ and $\Gamma_{\gamma^*}$ is the plasmon width,
\begin{equation}
\Gamma_{\gamma^*}=\frac{\left|d_{\mu}\right|^2 K^4}{6 \pi \omega}\left(1-\frac{m_N^2}{K^2}\right)^2\left(1+2 \frac{m_N^2}{K^2}\right) \theta\left(K-m_N\right)
\end{equation}
The plasma characteristics for a red-giant core before helium ignition are taken from Table D.1 in~\cite{Raffelt:1996wa}. 
\begin{equation}
\omega_p=18 \mathrm{keV},\, T_\gamma=8.6 \mathrm{keV},\, n_e=3 \times 10^{29} \mathrm{~cm}^{-3}
\end{equation}
The bound on $m_N$ and $d_{\mu}$ can be obtained by using the energy loss function $Q$ with $m_N=0,\;d_{\mu}<1.1\times 10^{-22}\mu_B$~\cite{Diaz:2019kim}.

\noindent {\it Effective Neutrino Number $ N_{\rm eff}$ Constraints.}
The sterile neutrinos bring impacts on evolution of the late Universe in two ways. On one side, they are always in equilibrium with the SM neutrinos, since the mixing between them is large enough. On the other side, the introduction of dipole interaction may affect the decoupling of the neutrino plasma, it depends on the relative strength between dipole and the SM electroweak interactions. According to ref.~\cite{Li:2022dkc}, the averaged dipole interaction rate is $\langle \sigma v n \rangle_{d} \approx 3.32 \alpha d_\mu^2 T^3$, while the SM electroweak interaction rate is at $\langle \sigma v n \rangle_{\rm{SM}}\approx G_F^2 T^5$. Thus the EW interaction would always be dominant, a rough estimation shows that the dipole interaction rate is just $1.3\%$ of the EW interaction rate at the neutrino decoupling temperature $\sim \mathcal{O}(\rm{MeV})$. Hence one could conclude that the sterile neutrino will keep in equilibrium with the SM neutrinos, and the neutrino plasma would decouple from the electromagnetic plasma at a comparable temperature as in standard cosmology. 

\begin{figure}
    \centering
    \includegraphics[width=0.45\textwidth]{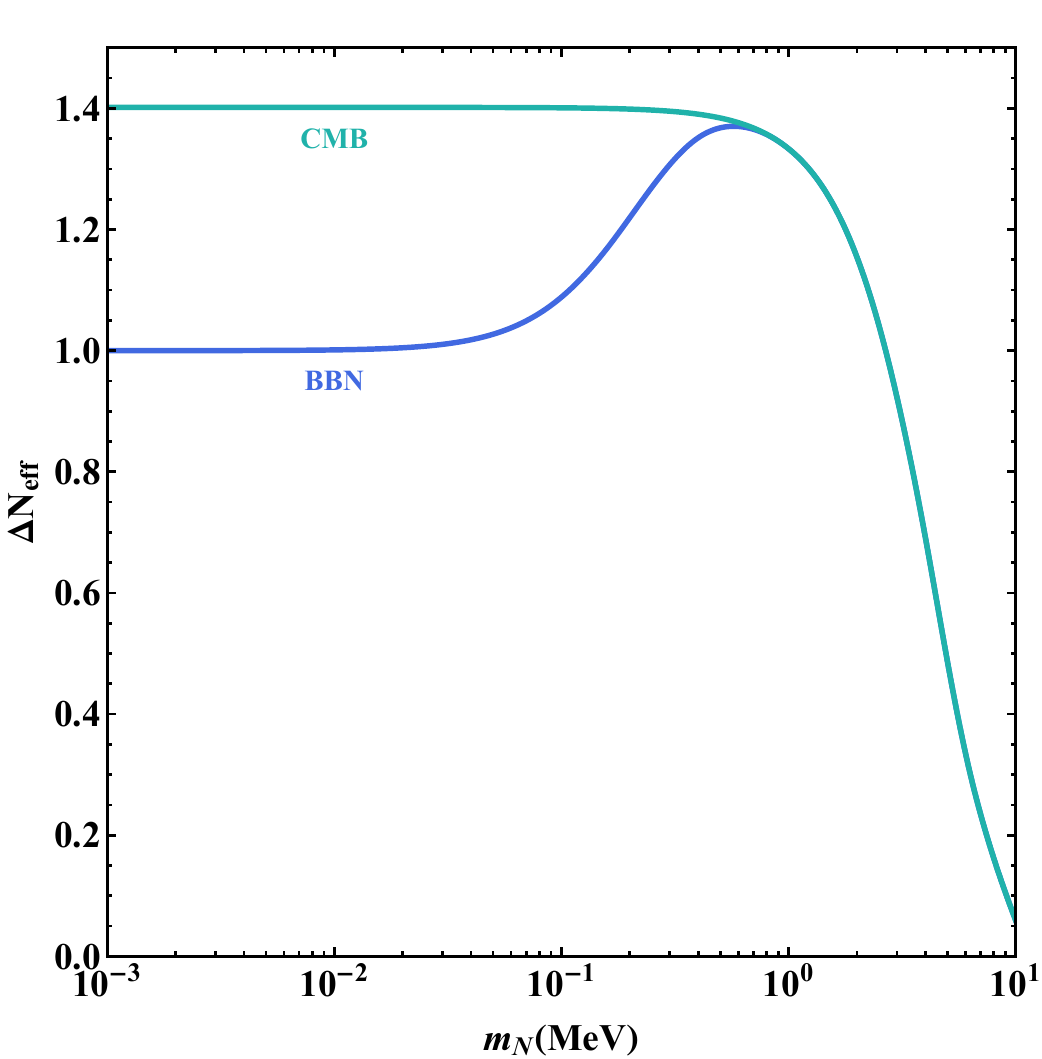}
    \caption{The distribution of $\Delta N_{\rm{eff}}$ with varied $m_N$, at the BBN and CMB epoch.}
    \label{fig:neff}
\end{figure}

We can evaluate $N_{\rm{eff}}$ as follows. The entropy conservation points out that $S= a^3 \left(\rho + p\right)/T = 2\pi^2/45 a^3 g^*_s T^3$ is constant. In the early universe when the neutrino plasma and electromagnetic plasma are in thermal equilibrium, they share the same temperature. While at $T<T_D$ the temperature ratio is given by
\begin{equation}
    \frac{T_\nu}{T_\gamma} = \left(\frac{g^*_{s:\nu,D}}{g^*_{s:\gamma,D}} \frac{g^*_{s:\gamma}}{g^*_{s:\nu}}\right)^{1/3},
\end{equation}
where the quantities with subscript $D$ reminds us that they are given at the neutrino plasma decoupling temperature $T_D$. The $N_{\rm{eff}}$ is fixed by comparing the total energy density
\begin{equation}
    \rho_R = \rho_\gamma \left[1+ \frac{7}{8} \left(\frac{T_\nu^0}{T_\gamma}\right)^4 N_{\rm{eff}}\right]
    \label{eq:rhoR}
\end{equation}
to the explicit energy density of the neutrino plasma~\cite{Boehm:2012gr}
\begin{equation}
    \rho_{\nu:N} = \rho_\gamma \frac{7}{8} \left(\frac{T_\nu}{T_\gamma}\right)^4 \left[N_\nu + \frac{g_N}{2} I(y_N) \right].
    \label{eq:rhonu}
\end{equation}
The $I(y_N)$ in Eq.~\ref{eq:rhonu} has a form of 
\begin{equation}
    I(y_N) = \frac{120}{7\pi^4} \int_y^\infty d\xi \frac{\xi^2 \sqrt{\xi^2 -y_N^2}}{e^\xi + 1},
\end{equation}
with $y_N\equiv m_N/T_\nu$.
The $T_\nu^0$ in Eq.~\ref{eq:rhoR} stands for the neutrino temperature in standard cosmology, while $T_\nu$ in Eq.~\ref{eq:rhonu} represents the temperature of neutrino plasma. We show the deviation $\Delta N_{\rm{eff}}$ in Figure \ref{fig:neff}, at BBN and CMB epoch respectively. Once sterile neutrinos become non-relativistic after decoupling, they can serve as a source of reheating for the neutrino plasma, analogous to the way electron-positron annihilation reheats the photon plasma. Consequently, the temperature of the neutrino plasma increases, which results in an increased value of $\Delta N_{\rm{eff}}$. The current limits on $\Delta N_{\rm eff}$ are set as $\Delta N_{\rm eff}<0.5$ at the BBN ~\cite{Blinov:2019gcj} and $\Delta N_{\rm eff}<0.28$ at the CMB~\cite{Planck:2018vyg}. In addition, 
the lifetime of $N$ is restricted to be less than 1 second to avoid significant modifications to the abundance of light elements in the BBN era~\cite{Bolton:2019pcu}. In our model, the dominant decay channel for $N$ is through dipole interactions, which allows us to translate this constraint on the lifetime of $N$ into constraints on $d_\mu$ and $m_N$.

\begin{figure}
    \centering
    \includegraphics[width=0.49\textwidth]{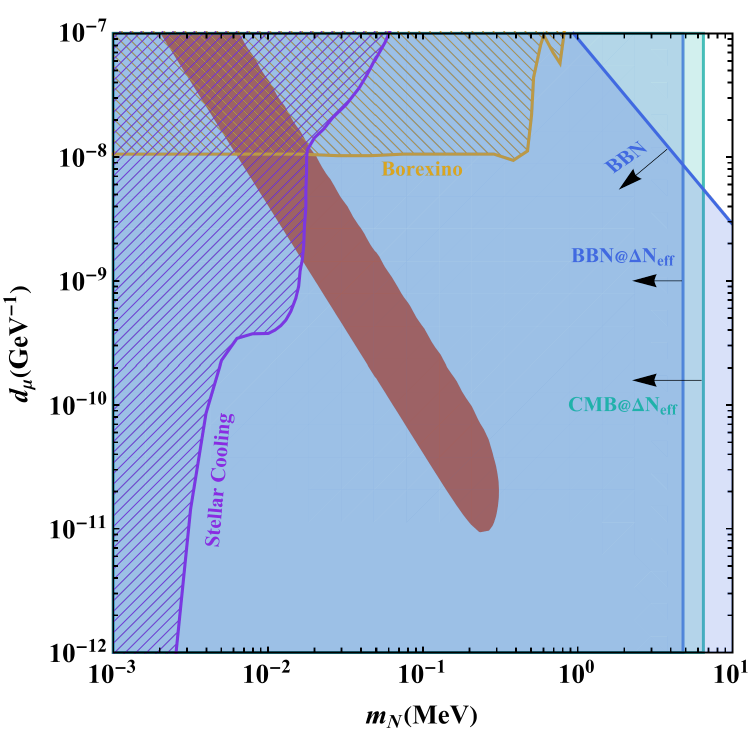}
    \caption{ The parameter space that produces more than one event in LHAASO experiment is shown in the plane of the dipole interaction strength $d_\mu$ versus sterile neutrino mass $m_N$ (red shaded region). Here we take $|U_{\mu 4}|=0.1$ in the calculation. For a smaller value of $|U_{\mu 4}|$, the allowed region would shrink. The blue and green shaded regions are excluded by BBN and CMB bounds. BBN bounds are obtained from constraints on $\Delta N_{\rm eff}$ and abundance of light elements respectively, while the CMB bound is solely derived from constraint on $\Delta N_{\rm eff}$. The purple and yellow hatched regions denote the limits from stellar cooling and Borexino.}
    \label{fig:widip}
\end{figure}

In Figure~\ref{fig:widip}, we show the allowed parameter space that can produce more than one energetic photon observed by LHAASO (red shaded region) and also plot the constraints mentioned above in the plane of the dipole interaction strength $d_\mu$ versus sterile neutrino mass $m_N$.  Our study assumes a value of $|U_{\mu 4}|=0.1$, which represents the maximum value permitted by the muon neutrino disappearance search conducted on MINOS~\cite{MINOS:2017cae}. For a smaller value of $|U_{\mu 4}|$, the corresponding region would shrink. From Figure~\ref{fig:widip}, we can see that the stellar cooling bound (purple hatched region) will become weak as the sterile neutrino mass increases. Both BBN (blue shaded region) and CMB (green shaded region) bounds on the effective neutrino number in Figure ~\ref{fig:neff} can entirely exclude the feasible parameter space for explaining the energetic photon events. The BBN bound derived from the restriction on the abundance of light elements offers a complementary excluding capability in addition to the constraints imposed by the effective neutrino number.
Our CMB/BBN limits extend the range of constraints on unstable more massive neutrinos, obtained in~\cite{Bolton:2019pcu} from other observational arguments. Besides, we also present the Borexino constraint from the measurement of the neutrino-electron scattering~\cite{Borexino:2017fbd} (yellow hatched region). It has set a limit on the effective dipole moment as $d_{\rm eff} < 2.8 \times 10^{-11} \mu_{\rm{B}}$, while a weaker limit is derived under the assumption of sterile neutrinos couple solely to $\nu_\mu$~\cite{Brdar:2020quo}.

\noindent {\it Other Possible Constraints.} For a long-lived sterile neutrino, such as $m_N \sim 100$ eV, the lifetime at rest of order $10^6$s so that its decays should take place in the period when electromagnetic energy release cannot restore Planck form of spectrum. Naive estimation, assuming that such electromagnetic energy release corresponds to half of $N$ energy density, which is 3/11 of photon energy density, and that this energy release leads to Bose-Einstein type distortion of CMB spectrum leads to a two-order of magnitude contradiction with the upper limit on such distortion. However, this statement needs more detailed analysis.
According to~\cite{zeldovich1969interaction}
and \cite{sunyaev1970interaction}
early energy release (at $z > 10^4$) leads to the heating of plasma electrons and their successive (inverse) Thomson scattering with CMB photons provides the formation of the Bose-Einstein spectrum with photon chemical potential determined by energy release. This point implies a special study in our case. Indeed, at $10^6$~s plasma temperature is 1 keV. $N$ gas temperature should be factor $(4/11)^{1/3}$ smaller, but still makes a dominant fraction of $N$ relativistic, and their decay is delayed by a factor of 10, so that they dominantly decay at $10^7$~s, when plasma and radiation temperature are about 300 eV (and maximum of Planck spectrum is near 1 keV). At this temperature, $N$ decay is not at rest, and the monochromatic photon distribution of two-body decay converts into the interval of energies $E \le 150$ eV. At such energy photons make energy transfer to electron $(E/m)E \le 3 \times 10^{-4}E$ in each collision and electron heating by such energy release becomes problematic. A rigorous analysis, which should involve a detailed study of the character of CMB spectrum distortions, should be undertaken in our case, but even without it, it is qualitatively evident that $N$ decays would contribute to the low frequency (Raleigh-Jeans) part of CMB spectrum, increasing it by a factor of $\delta \epsilon/\epsilon$, where $\delta \epsilon \sim E_N n_N$ and $\epsilon$ is CMB energy density in the period of decay. These calculations should be done more rigorously to foresee possible probes for low-energy new physics in the searches by precision cosmology for CMB spectral distortions (see e.g., \cite{Khlopov:2022zqr}). Remind that the results of this section correspond to the range of neutrino mass~ 100 eV.

\section{Conclusion}\label{conclusion}

Confrontation of multimessenger astronomical observations of GRB221009A with multimessenger astrophysical and cosmological probes makes an incompatible interpretation of these observations with the sterile neutrino hypothesis. Even if the ultra high energy LHAASO event is in occasional coincidence with GRB221009A and observation of this gamma-ray burst doesn't imply the involvement of new physics, our analysis has to lead to a new constraint on parameters of sterile neutrino physics and revealed a possible new type of deviations from the standard Big Bang scenario originated from effects of new physics at the low energy scale. A possibility of such effects may be useful for astrophysical studies and stimulating searches for a new type of CMB spectral distortions as cosmological probes for new physics. 

\begin{acknowledgments}
 This work by S.G., L.W. and B.Z. was supported by the National Natural Science Foundation of China under grants No. 12005180, 12275134 and 12275232, by the Natural Science Foundation of Shandong Province under Grants No. ZR2020QA083, ZR2022QA026, and by the Project of Shandong Province Higher Educational Science and Technology Program under Grants No. 2019KJJ007. The research by M.K. was financially supported by Southern Federal University, 2020 Project VnGr/2020-03-IF.

\end{acknowledgments}

\bibliography{refs}

\end{document}